\documentclass{article}
\pdfoutput=1
\usepackage{jheppub}
\usepackage{dsfont}
\usepackage{amssymb,amsmath}
\usepackage{epsfig}
\usepackage{epstopdf}
\usepackage{latexsym}
\usepackage{graphicx}
\usepackage{subfigure}
\usepackage{booktabs}
\usepackage{bbm}
\usepackage{tikz}
\pdfoutput=1
\makeatletter
\def\@fpheader{\relax}
\makeatother

\def\cM{{\cal M}}

\def\cO{{\cal O}}

\def\cU{{\cal U}}

\def\cM{\mathcal{M}}

\def\cO{\mathcal{O}}

\def\cU{\mathcal{U}}

\def\beq{\begin{eqnarray}}
\def\eeq{\end{eqnarray}}

\def\ie{{\it i.e.}}

\def\be{\begin{equation}}
\def\ee{\end{equation}}
\def\bea{\begin{eqnarray}}
\def\eea{\end{eqnarray}}

\def\cM{\mathcal{M}}

\def\cO{\mathcal{O}}

\def\cU{\mathcal{U}}

\definecolor{AV}{rgb}{0.65,0.0,0}
\definecolor{AK}{rgb}{0,0,0.65}

\title{Oscillating Shells and Oscillating Balls in AdS}
\author{Avik Banerjee$^{a,c}$, Arnab Kundu$^{a,c}$, Pratik Roy$^{b,c}$, Amitabh Virmani$^{b,c}$}
\affiliation{$^a$Theory Division, Saha Institute of Nuclear Physics, \\ 1/AF Bidhannagar, Kolkata 700064, India.}
\affiliation{$^b$Institute of Physics, \\ Sachivalaya Marg, Bhubaneswar 751005, Odisha, India.}
\affiliation{$^c$Homi Bhaba National Institute, \\ Training School Complex, Anushakti Nagar, Mumbai 400085, India.}
\emailAdd{avik.banerjee, arnab.kundu@saha.ac.in, \\ pratik, virmani@iopb.res.in}

\abstract{It has recently been reported that  certain thin timelike shells undergo oscillatory motion in AdS.  In this paper, we compute two-point function of a probe field in the geodesic approximation in such an oscillating shell background. We confirm that the two-point function exhibits an oscillatory behaviour following the motion of the shell. We  show that similar oscillatory dynamics is possible when the perfect fluid on the shell has a  polytropic equation of state. Moreover, we show that certain ball like configurations in AdS also exhibit oscillatory motion and comment on how such a solution can be smoothly matched to an appropriate exterior solution. We also demonstrate that the weak energy condition is satisfied for these oscillatory configurations.}

\begin{document}

\maketitle

\section{Introduction and summary}

In recent times the study of thermalisation in closed quantum systems has received a surge of activity, see {\it e.g.}~\cite{Polkovnikov:2010yn} for a review with more references. In general, a quantum system perturbed out of equilibrium  decoheres and proceeds towards ergodicity. On a large enough time-scale, the system thermalises and  is described by a mixed density matrix. However, contrary to this expectation, there can be situations where a quantum system dynamically reconstructs the initial state and keeps repeating this evolution, with or without damping. This phenomenon is termed quantum revival.

In the context of AdS/CFT correspondence, such possible revival configurations presumably correspond to periodic or quasi-periodic dynamics resulting from gravity in an asymptotically AdS space. In this paper, we discuss such configurations in two contexts, namely, thin shells and solid balls especially when they oscillate. Oscillatory motion in gravitational dynamics is not new. In global  AdS space, a transient oscillatory motion of a thick shell has been reported in~\cite{Bizon:2011gg, Buchel:2012uh}. The thick shell leads to a collapse situation, forming a black hole at late times. Non-transient or exactly periodic oscillatory configurations in AdS space have been explored in~\cite{Dias:2011ss, Maliborski:2013jca, Horowitz:2014hja, Bizon:2015pfa, Mas:2015dra, daSilva:2016nah}. A connection with quantum revivals has also been proposed.  
Similar periodic configurations are also known to arise in other closely related set-ups~\cite{Gao:2008jy, Delsate:2014iia, Rocha:2015tda, Cardoso:2016wcr, Brito:2016xvw}.

It becomes clear from these studies that we need two ingredients for oscillatory dynamics in AdS. Firstly, we need to work in  global AdS, and secondly, we need a non-vanishing pressure (or another repulsive force) to sustain oscillations. The necessity of a non-vanishing pressure is intuitive. To have an oscillation, one needs an interaction that competes with the attraction of gravity. In earlier works, {\it e.g.}~in~\cite{Mas:2015dra, Taanila:2015sda}, oscillatory shells have been explored, where the shell matter is described by a perfect fluid with a linear equation of state. In the current paper, we also consider a polytropic equation of state for the shell dynamics, and a non-vanishing pressure for the ball dynamics. In both cases, we conclude that for a range of allowed parameter space, one obtains oscillatory motion. Furthermore, reasonable energy conditions, such as the weak and null energy conditions are obeyed by these configurations.

While {\it a priori}  there is no reason to rule out such dynamics, it remains unclear to us what the precise dual field theory descriptions are. One such possibility is certainly the {\it quantum revivals} that have already been pointed out in the literature~\cite{daSilva:2016nah}. We only list a few features here, and not attempt to elaborate on the identification. First, it is clear that one point functions are all thermal as seen from the AdS  boundary. Non-local observables, however, do penetrate and capture the dynamical aspects of the geometry. Towards this we explicitly calculate a two-point function in the geodesic approximation in oscillatory shell backgrounds and demonstrate that the shell oscillation simply gets mapped to oscillations of the correlation function, provided the two points are sufficiently separated at the boundary.

We expect a similar behaviour to appear in the oscillating ball dynamics, though, this calculation is technically more involved. The technical complication for the ball dynamics arises from a non-vanishing pressure. The ball itself is described by a simple FRW geometry, and due to pressure matter leaks outside. The outside is therefore a Tolman-Oppenheimer-Volkoff (TOV) type solution in AdS. This TOV geometry needs to be matched onto an AdS Schwarzschild geometry. Thus, a two-point function in the dual field theory has essentially three characteristic length-scales. The short-distance behaviour of the correlator is purely thermal. The intermediate-distance behaviour of the correlator  is determined by a geodesic penetrating into the TOV region of the spacetime. Finally, the long-distance behaviour of the correlator is dynamical since the corresponding geodesic probes the oscillating FRW region. Thus, the UV modes of the field theory have a thermal behaviour, which crosses over to a dynamical behaviour towards the IR. This qualitative picture is in accordance with the {\it top-down} thermalisation picture of~\cite{Balasubramanian:2010ce, Balasubramanian:2011ur, Garfinkle:2011hm, Garfinkle:2011tc} in the context of AdS/CFT correspondence.

Another intriguing feature of the oscillatory configurations is that the dynamics is confined between two radial scales.  One does not immediately arrive at such a configuration with a natural choice of boundary and initial conditions at the boundary of AdS. Thus, while the presence of oscillations is rather ubiquitous, our analysis does not shed light on how one prepares this state from the perspective of the boundary theory. However, given the results of~\cite{daSilva:2016nah, Mandal:2016cdw}, where a more direct numerical study  exhibits similar periodic or quasi-periodic dynamics arising from a set of initial and boundary conditions, we view the above shortcoming as a limitation of our approach.

Given the existence of the oscillatory dynamics, there are various avenues to explore further, for example, how  additional parameters affect the oscillatory configurations? In particular, introducing a charge is potentially interesting since it can compete with gravitational attraction. Perhaps with a non-vanishing charge, one can obtain oscillating solutions in low pressure situations. We leave this for future investigations.

The rest of the paper is organised as follows. In section \ref{osc_shells} we discuss the basic framework of junction conditions and the details of the shell dynamics, including the study of two point function in the geodesics approximation in oscillating shell backgrounds. In section \ref{osc_balls} we discuss oscillating FRW balls. Section \ref{energy_conditions} is devoted to a discussion of the various energy conditions for the oscillating shell and oscillating ball configurations. Finally, certain details on matching the FRW ball to a TOV-type solution are discussed in appendix \ref{OS_p_AdS}.

\section{The oscillating shells}

\label{osc_shells}

In this section we discuss oscillating shell configurations.  In section \ref{sec:shell} we start with a brief review of the analysis of \cite{Mas:2015dra} and make some further observations. In section \ref{geodesics} we compute the equal time two-point function in the geodesic approximation in oscillating shell backgrounds.

\subsection{Shell dynamics}
\label{sec:shell}

We begin with the formalism to discuss the motion of the shell. The same formalism will be useful later in studying the dynamics of a ball in the spirit of the Oppenheimer-Snyder model.

Consider a spherically symmetric thin shell, evolving in a $(d+1)$-dimensional background spacetime $\mathcal{M}$. The shell divides the entire spacetime in two regions: an interior (empty AdS) denoted by $\mathcal{M_{-}}$ and an exterior (AdS Schwarzschild) denoted by $\mathcal{M_{+}}$. The line elements in the two regions are given by
\begin{equation}
ds_{\pm}^2= -f_{\pm} (r)dt_{\pm}^2 + f_{\pm }^{-1}(r)dr^2 + r^2 d\Omega_{d-1}^2,
\end{equation}
where $f_-(r)= 1+ r^2 $ and $f_+(r)= 1+ r^2 - \frac{m}{r^{d-2}} $. Here, we have set the AdS length to unity and $m$ is the mass parameter (proportional to the ADM mass) of the system.

The radial coordinate $r$ is continuous across the shell, ensuring that the area of the $(d-1)$-spheres agree on the two sides of the shell. In brief, we choose the following coordinate patches: $\cU_+ \equiv \{t_{+}, r, \theta_1, \ldots \theta_{d-1}\} \equiv \{x_+^\mu\}$ on $\cM_{+}$, and $\cU_- \equiv \{t_{-}, r, \theta_1, \ldots \theta_{d-1}\} \equiv \{x_-^\mu\} $ on $\cM_{-}$. Here $\mu$ ranges over all space-time directions. Clearly, Einstein equations (via the junction conditions)  impose non-trivial boundary conditions on $\cU_+ \, \cap \, \cU_-$, thereby determining the entire manifold covered by $\cU_+ \cup \, \cU_-$.

We can choose an independent set of coordinates on the shell worldvolume $$ \ \cU_{\rm shell} \equiv \left \{ \tau , \theta_1 , ....,\theta_{d-1} \right \} \equiv \{ y^a \}.$$ In writing this equation, we have chosen a trivial embedding along the angular directions by making use of the spherical symmetry of the problem. The coordinate $\tau$ is chosen to be the proper time of a co-moving observer on the shell. The basis vectors on the tangent space of the shell at any point can be pushed forward to spacetime vectors: $e_{\rm space-time} \equiv \varphi_*\left( e_{\rm shell } \right)$. In  explicit coordinates, this map takes the form $\partial_{a}= \frac{\partial x^{\mu}}{\partial y^{a}}\partial_\mu$.

Let the position of the shell be specified by 
\begin{equation}
r=r_{\rm s}(\tau)\ ,\hspace{5mm} t_{\pm}=t_{\pm, \rm s}(\tau).
\end{equation} 
Then we get
\begin{eqnarray}
&& \partial_{\tau} = u^{\mu}\partial_{\mu} = \dot{t}_{\pm, \rm s} \ \partial_{t_{\pm}} + \dot{r_{\rm s}} \partial_{r} , \\
&& \partial_{\theta_{i}} = \delta_{\theta_{i}}^{\mu}\partial_{\mu}, \quad i = 1, \ldots, (d-1) .
\end{eqnarray}
Here, the overhead dot denotes derivative w.r.t.~$\tau$, and $u^{\mu}$ is the four velocity of the shell.  The four velocity is canonically normalised, $u^{\mu}u_{\mu}=-1$, which yields,
\begin{equation}
\dot{t}_{\pm, \rm s } = \frac{\sqrt{f_{\pm}(r_{\rm s})+\dot{r_{\rm s}}^2}}{f_{\pm}(r_{\rm s})} =:  \frac{\beta_{\pm}}{f_{\pm}(r_{\rm s})}. \label{tdotshell}
\end{equation}
Since the derivatives  $\dot{t}_{\pm}$ do not match at the location of the shell, $t_+$ is not continuously related to $t_-$. This will be carefully taken into account when we discuss spacelike geodesics crossing the shell in the next subsection.

The induced metric on the shell is:
\begin{equation}
ds_{\rm shell}^2= h_{ab}dy^a dy^b= -d\tau^2 + r_{\rm s}^2 d\Omega_{d-1}^2. \label{shellinduced}
\end{equation}
The unit normalised vector normal  to the shell in $\pm$ coordinates is
\begin{eqnarray}
n_{\mu,\pm} = \left(- \dot{r_{\rm s}}, \dot{t}_{\pm,{\rm s}},0,0 \right) \quad \implies \quad  n^{\mu}_{\pm}= \left(f^{-1}_{\pm}(r_{\rm s})\dot{r}_{\rm s}, f_{\pm}(r_{\rm s})\dot{t}_{\pm,{\rm s}},0,0 \right). \label{normals}
\end{eqnarray}
It satisfies $u^{\mu}n_{\mu,\pm}=0$ and $n_{\pm}^{\mu}n_{\mu,\pm}=1$. An overall (positive) sign choice has been made in writing the above normal vector, so it points from $\cM_-$ to $\cM_+$.

Einstein equations become a set of matching conditions on $\cU_+ \cap \cU_-$. These are known as the Israel junction conditions~\cite{Poisson}. For writing down these conditions, we need to evaluate the extrinsic curvature and also assign a stress-tensor to the thin-shell matter field. The extrinsic curvature, defined as $K_{ab}= e ^{\mu}_{\, \, \, a} e^{\nu}_{\, \, \, b} \nabla_{\mu}n_{\nu}$, has the following non-zero components,
\begin{eqnarray}
&& K_{\tau \tau,\pm} = -\frac{\dot{\beta_{\pm}}}{\dot{r}_{\rm s}}, \quad K_{\theta_1 \theta_1,\pm} = \beta_{\pm}r_{\rm s} , \\
&& K_{\theta_i\theta_i,\pm} = \left (K_{\theta_1 \theta_1,\pm} \right) \frac{h_{\theta_i \theta_i}}{r_{\rm s}^2}, \quad i \not = 1.
\end{eqnarray}
Equivalently,
\bea
&& K^{\tau}_{\, \, \, \tau,\pm} = \frac{\dot{\beta_{\pm}}}{\dot{r}_{\rm s}}, \quad K^{\theta_i}_{\, \, \, \, \theta_{i,\pm}} = \frac{\beta_{\pm}}{r_{\rm s}}. \label{extrinsic}
\eea%
For simplicity, we can take the stress-tensor of the thin-shell to be of the perfect fluid form,
\begin{equation}
S^a_{\, \, \, b} = {\rm diag}( -\sigma,\underbrace{p,p, \ldots }_{(d-1) \, {\rm terms}} ), \label{shellmatter}
\end{equation} 
where $\sigma$ and $p$ are the energy density and the pressure of the corresponding matter on the shell, related via a suitable equation of state.

The two Israel junction conditions are (i) continuity of metric across the shell, and (ii) jump in the extrinsic curvature is related to the stress-tensor of the thin-shell,
\begin{equation}
\left[K_{ab} \right]- h_{ab}\left[K \right]= - \kappa S_{ab}, \quad {\rm or} \quad  \left[K^a_{\, \, \, b}\right] = - \kappa \left(S^a_{\, \, \, b} - \frac{\delta^a_{\, \, \, b} S}{d-1}\right), \label{IsraelJC}
\end{equation}
where $\kappa = 8\pi G_{d}$, $K \equiv h^{ab} K_{ab}$ and $S \equiv h^{ab} S_{ab}$ are the traces of the corresponding tensors. The bracket, denoted by $[]$ represents the jump  from $\cM_-$ to $\cM_+$
\be
\left[ \cO\right] \equiv \cO_+ - \cO_-
\ee for some field $\cO$.  This definition is tied to our convention of choosing the direction of the normal vector in (\ref{normals}).

Together with \eqref{extrinsic} and  (\ref{shellmatter}), the junction conditions (\ref{IsraelJC}) become
\begin{eqnarray}
\frac{[\beta]}{r_{\rm s}} &=& -\frac{\kappa\sigma}{d-1},  \label{israelexplicit} \\
\frac{[\dot{\beta}]}{\dot{r}_{\rm s}} &=& \kappa \left(p + \sigma \frac{d-2}{d-1} \right).  \label{israelexplicit2}
\end{eqnarray}
Since, $f_+ \le f_-$, we have $\beta_{+} \le \beta_{-} $, and by virtue of (\ref{israelexplicit}), we conclude $\sigma \ge 0$. The inequality here is saturated for the trivial junction where the extrinsic curvature has no jump, and the shell does not exist.

To make further progress, one needs to input an equation of state. A sufficiently general choice is the polytropic equation of state: $ p = \frac{\alpha}{d-1} \sigma^\gamma $, where $\gamma$ is the polytropic exponent. The overall constant $\alpha$ fixes the normalization of {\it e.g.}~the trace of the shell energy-momentum tensor. With a polytropic equation of state, equation (\ref{israelexplicit2}) takes the form,
\be 
\frac{[\dot{\beta}]}{\dot{r}_{\rm s}} = - \frac{[\beta]}{r_{\rm s} } \left (\alpha \sigma^{\gamma -1 } + d - 2 \right) \label{betaeq},
 \ee
which can be integrated using \eqref{israelexplicit} to yield,
\bea
- \left[\beta \right]  =  \left[- \frac{ \left( d-1\right)^{\gamma -2}}{\kappa^{\gamma -1 }}\ \alpha \ r_{\rm s}^{1 - \gamma  } + M r_{\rm s}^{\left( d-2 \right) \left( \gamma -1 \right) }\right]^{\frac{1}{1-\gamma}}, \quad \gamma \in {\mathbb Z} \setminus \{ 1 \}\ ,  \label{1stint}
\end{eqnarray}
where $M$ is a constant of motion.

It is also possible to obtain analytical solutions   for equation \eqref{betaeq} with non-integer values\footnote{This is certainly of physical importance, see \emph{e.g.},~\cite{Chandra}.} of $\gamma$, however, those seem valid case-by-case and we were not able to obtain one compact expression for all possible values of $\gamma$. 
The special case of $\gamma =1$ can be worked out separately, yielding,
\begin{eqnarray}
\left[ \beta \right] = -  M r_{\rm s}^{2 - d - \alpha}, \label{betasimple}
\end{eqnarray}
where $M$ is an integration constant\footnote{For $\sigma$ to be positive, the constant $M$ in equation \eqref{betasimple} needs to be positive, cf.~\eqref{israelexplicit}. For the polytropic equation of state, the relation between the integration constant $M$ and $\sigma$ is not direct. Since a physical interpretation of $M$ is not transparent, one can consider both positive and negative values of $M$ for the polytropic equation of state. In this paper we only consider  $M>0$.}. 
In this case we have a linear equation of state $p = \frac{\alpha}{d-1} \sigma$. There are two  cases of  special interest, $\alpha = 0$  and $\alpha = 1$. $\alpha = 0$  corresponds to pressure-less dust, and $\alpha = 1$ corresponds to conformal matter for which the trace of the energy-momentum tensor vanishes.

The total energy of the shell can be defined by \be E = \sigma \Omega_{d-1} r_{\rm s}^{d-1}. \label{energy}\ee In general, the energy so defined is clearly not conserved as $r_s$ and $\sigma$ change as the shell moves. However, in the pressure-less case, 
using \eqref{israelexplicit} one sees that  $E$ is a  constant of motion related to $M$ by a proportionality factor.

We can recast the equations of motion of the shell as the motion of a particle in an effective potential. This is achieved by  substituting the definition 
$\beta_{\pm} = \sqrt{f_{\pm}(r_{\rm s})+\dot{r_{\rm s}}^2}$ in equation (\ref{israelexplicit}). After some simplification we get, 
\begin{eqnarray}
&& \dot{r}_{\rm s}^2 + V_{\rm eff} \left(r_{\rm s} \right) = 0 , \label{shelldyn1} \\
&& V_{\rm eff} \left(r_{\rm s} \right) = f_{-}\left(r_{\rm s}\right) - \frac{(d-1)^2}{4\sigma^2 r_{\rm s}^2} \bigg[f_{-} \left(r_{\rm s}\right) - f_+ \left(r_{\rm s}\right) + \frac{\sigma^2 r_{\rm s}^2}{(d-1)^2}\bigg]^2 \label{shelldyn2}.
\end{eqnarray}
In practice, one uses \eqref{israelexplicit}  to write $\sigma$ as,
\be
 \sigma = - \frac{\left[ \beta \right] \left(d -1 \right) }{\kappa r_{\rm s}}, 
\ee
and in turn uses (\ref{1stint}) to substitute for $[\beta]$ to obtain $\sigma$ in terms of other parameters.  Substituting such an expression in \eqref{shelldyn2}, gives an equation for 
 the dynamics of the shell  in terms of the paramters $\{d, \alpha, \gamma, \kappa, m, M\}$. Of these, $\kappa$ can be set to unity by an appropriate choice a  units. Therefore, the physics depends on parameters $\{d, \alpha, \gamma, m, M\}$.
\begin{figure}[t]
\begin{center}
{\includegraphics[width=0.55\textwidth]{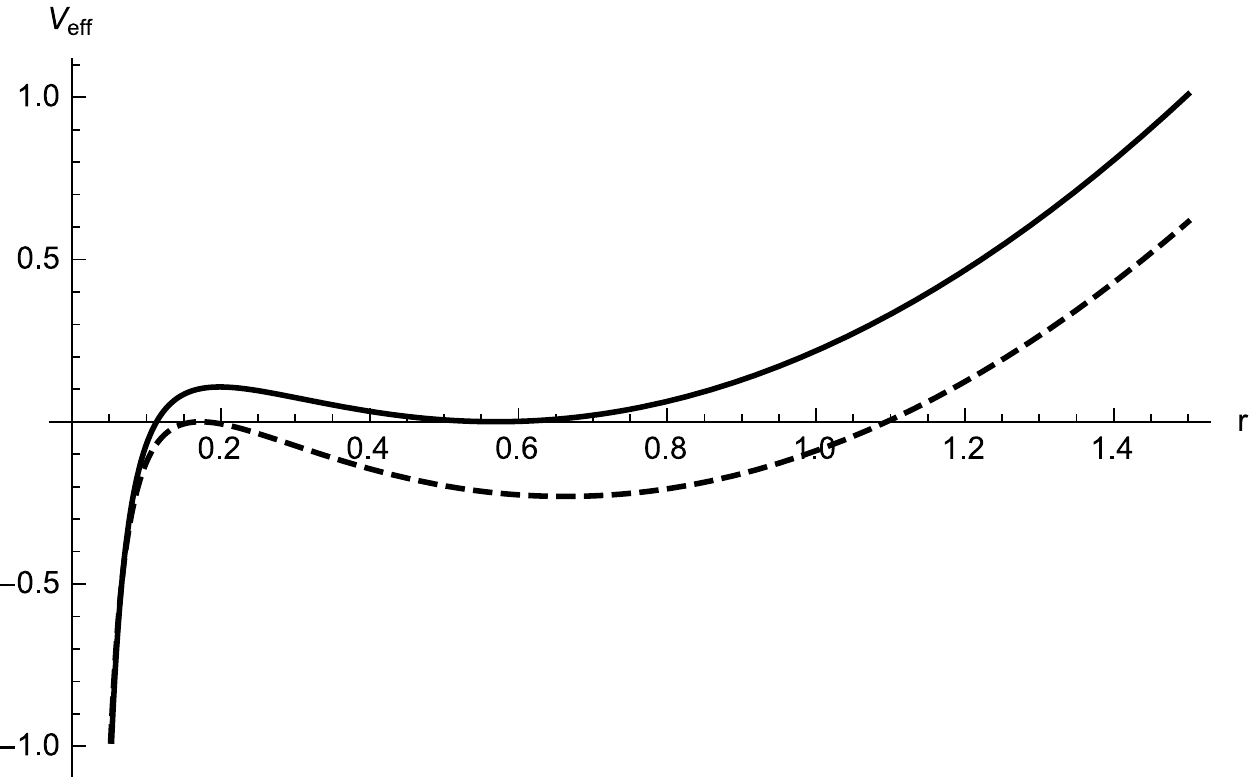}}
\caption{\sf The two limiting cases within which  oscillatory shell motion exists. We have chosen $\alpha = 0.3$, $d=3$, $m=0.1$, $M_{\rm up} / m = 0.38$ (the top solid curve) and $M_{\rm low} / m =0.35$ (the bottom dashed curve).} \label{oscipot}
\end{center}
\end{figure}

Note that
for $\gamma \not =1$, the two terms in (\ref{1stint}) compete with each other, and define a natural scale for the dynamics,
\begin{eqnarray}
\cO \left( \frac{ \left( d-1\right)^{\gamma -2}}{\kappa^{\gamma -1 }}\ \alpha \ r_{\rm cross}^{1 - \gamma  }  \right) = \cO \left( M r_{\rm cross}^{\left( d-2 \right) \left( \gamma -1 \right) } \right), 
\end{eqnarray}
where $r_{\rm cross}$ denotes the crossover scale which connects two different dynamical regimes, described respectively by an $r_{\rm s}^{1-\gamma}$ potential and the inverse of it. In general, with various possibilities, the full dynamics is likely to be very rich and worth exploring. We focus only on certain sub-classes in this paper.

Let us start by briefly reviewing the oscillatory solutions that are already discussed in \cite{Mas:2015dra}. This corresponds to setting $\gamma=1$. The effective potential can be rewritten as,
\begin{eqnarray}
V_{\rm eff} = 1 + r_{\rm s}^2 - \frac{m^2}{4M^2} r_{\rm s}^{2\alpha} - \frac{m}{2} r_{\rm s}^{2 - d} - \frac{M^2}{4} r_{\rm s}^{- 2(\alpha + d - 2)},
\end{eqnarray}
where $M$ is now the constant appearing in the first integral of motion in (\ref{betasimple}). To find oscillatory shell dynamics, one can proceed as follows. 

We impose $V_{\rm eff} = 0$ and $\partial_{r_{\rm s}} V_{\rm eff} = 0$, to find algebraic solutions characterized by \be 
\left\{m\left( d, \alpha, r_{\rm s} \right), M\left( d, \alpha, r_{\rm s} \right) \right\}.\ee These values can be viewed  as special cases, when two roots of the effective potential coalesce. 
See figure \ref{oscipot}.
Evidently, if this is a local minimum, and the effective potential can be lowered by tuning other parameters in the system, oscillatory shell dynamics will ensue. 
%
%
%
Explicit expressions for $m\left( d, \alpha, r_{\rm s} \right)$ and $M\left( d, \alpha, r_{\rm s} \right)$ are given in reference \cite{Mas:2015dra}. For a fixed value of mass $m_*$ less than a maximum value,
\be
m_* \le m_{\rm max}(d,\alpha),
\ee 
the equation \be
m_* = m\left( d, \alpha, r_{\rm s} \right),
\ee
yields two roots of $r_{\rm s}$, denoted by $r_{\rm up}$ and $r_{\rm low}$. The function $M\left( d, \alpha, r_{\rm s} \right)$ evaluated at these two roots yield two values of $M$, denoted by $M_{\rm up}$ and $M_{\rm low}$.  Choosing a value of $M$ such that 
\be 
M_{\rm low} \le M \le M_{\rm up},
\ee for a suitably fixed value of $m$, the shell undergoes oscillatory motion. 
The function $m_{\rm max}(d,\alpha)$ is such that for fixed $m$ the oscillatory solutions exist only beyond a critical non-zero value $\alpha = \alpha_{\rm crit}$. For $d=4$ we have explicitly checked that once we choose an $\alpha > \alpha_{\rm crit}$, the range $\left(M_{\rm up} - M_{\rm low} \right)$ increases with increasing $\alpha$. For $\alpha < \alpha_{\rm crit}$ only collapsing solutions exist. 

One can also see that as $\alpha$ approaches one, the maximum mass for which oscillating solutions exist, $m_{\rm max},$ increases without bounds \cite{Mas:2015dra}. As $m$ diverges, the upper turning point, $r_{\rm up}$, diverges with it. This means that one can tune $M$ such that in the limit that the shell is made of conformal matter and is collapsing from infinity it can develop oscillations.  

\begin{figure}[t]
\begin{center}
{\includegraphics[width=0.48\textwidth]{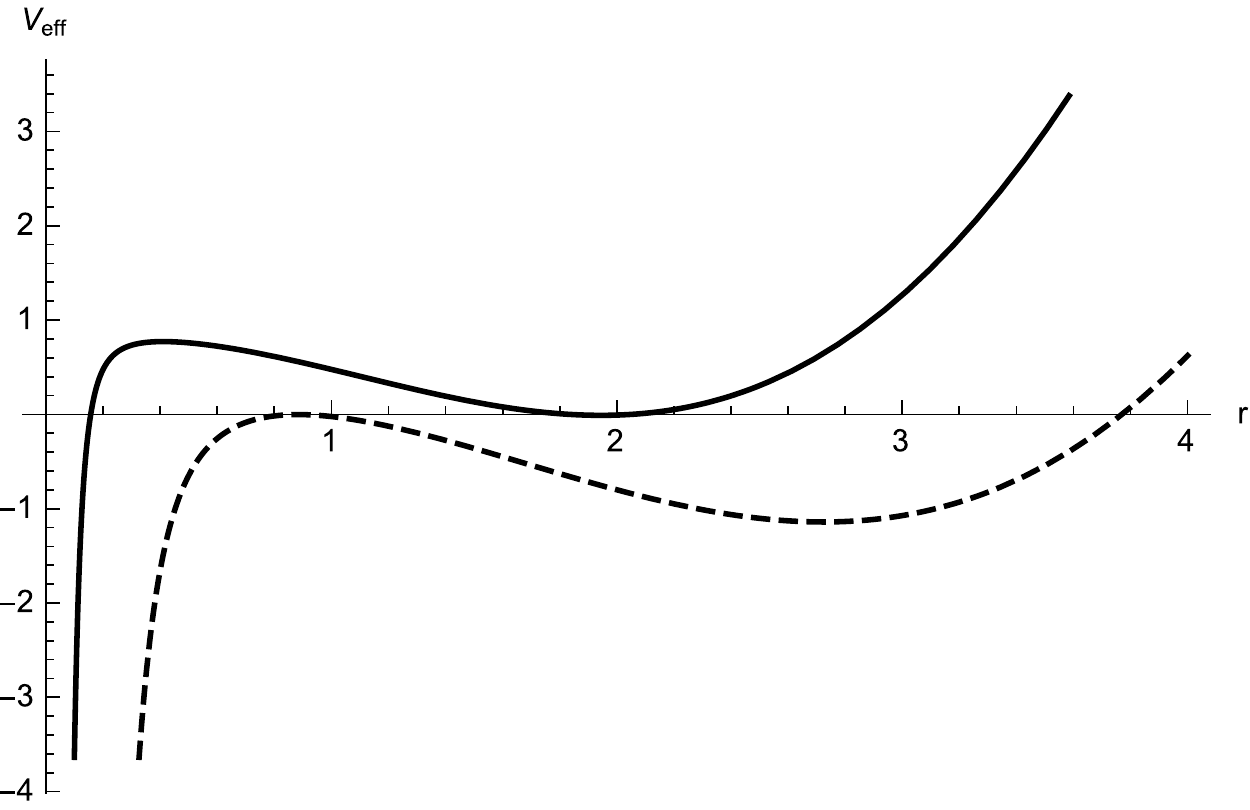}}
{\includegraphics[width=0.48\textwidth]{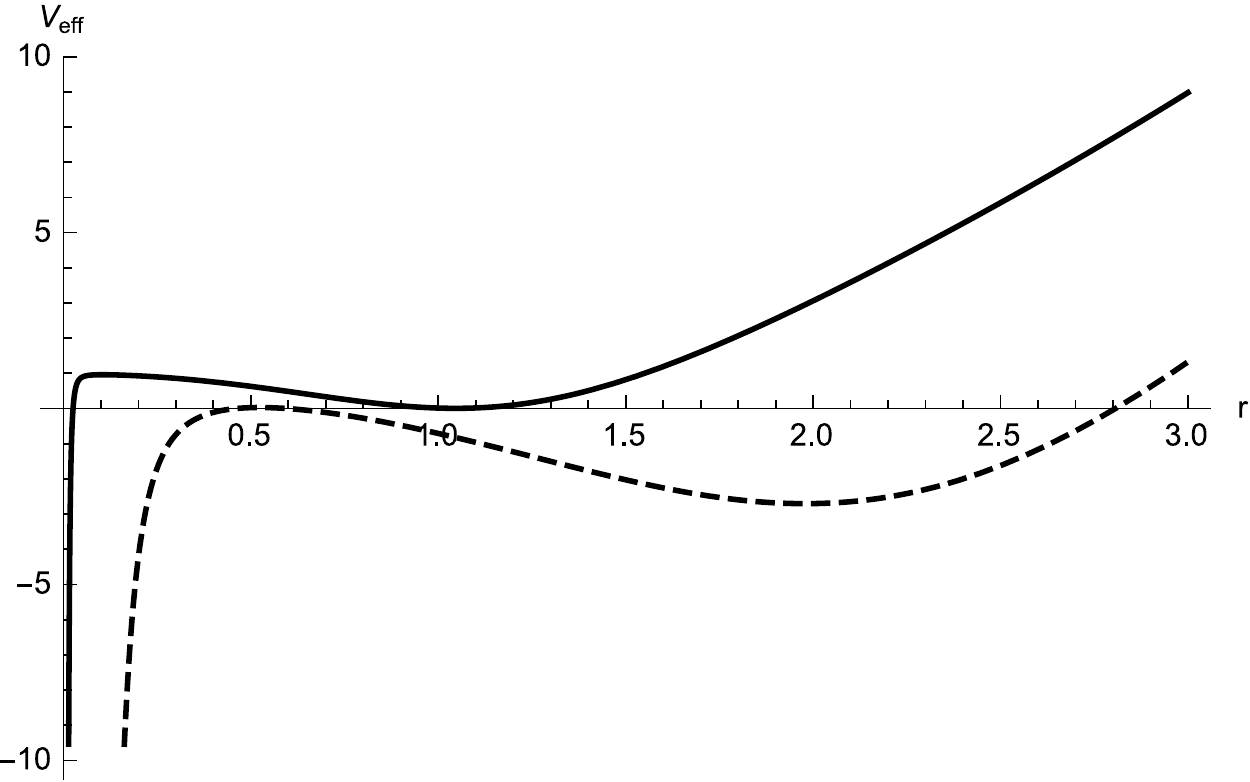}}
\caption{\sf 
The limiting cases within which oscillatory shell motion exists.
The plot on the left corresponds to $\gamma=2$, $\alpha = -  0.4$,  $m=0.94$, $M=0.006$ (dashed) and $m = 0.1$, $M=0.012$ (solid). The plot on the right corresponds to $\gamma=3$, $\alpha = - 0.05$, $m=0.5$, $M=0.002$ (dashed) and $m=0.004$, $M=0.026$ (solid). The two plots indicate that increasing $\gamma$ brings the two limiting cases closer to each other.} \label{VeffGamma}
\end{center}
\end{figure}

Let us comment on the typicality of such oscillatory configurations with a non-trivial polytropic exponent. In principle, the above analysis can be carried out for any value of $\gamma$. However, we only discuss explicitly the cases with $\gamma = 2$ and $\gamma=3$, which is perhaps sufficient for the generic story. The algebraic expressions associated with this analysis are fairly involved and we refrain from presenting them explicitly. Instead, we summarise the generic finding in figure \ref{VeffGamma} in terms of the features of the effective potential. The main features are as follows.

In producing figure \ref{VeffGamma}, we have chosen a negative value of $\alpha$ for both $\gamma =2$ and $\gamma=3$. 
It can be easily seen that as far as satisfying a reasonable energy condition is concerned, negative values for $\alpha$ are allowed. For example, ensuring weak energy condition requires $\sigma \ge 0$ and $\sigma + p \ge 0$. For positive $\sigma$ and for $\gamma = 1$, the weak energy condition only requires $ \alpha \ge - (d-1)$. As another example, for $\gamma=3$ weak energy condition only requires $\alpha \ge - \frac{d-1}{\sigma^2}$, which leaves a window for choosing a negative value of $\alpha$. For $\gamma=2,3$ we have also observed that the upper turning point goes to infinity as $M$ approaches zero.

Finally, for general integer values of $\gamma$, the oscillatory regime can be characterized by a $3$-tuple: $\left(\alpha, M, m \right)$. For a fixed value of $\alpha$, both $M$ and $m$ need to be tuned to obtain the potential well. In figure \ref{VeffGamma}, we have shown the corresponding extremal cases, by tuning both $M$ and $m$ to the respective values quoted in the figure caption. It is noteworthy that we have not found an oscillatory configuration along the  $\gamma < 0$ branch, assuming that both $m > 0$ and $M >0$.

To add further support to the existence of oscillatory configurations, we have also explored a few non-integer values of $\gamma$. In particular, we here comment on the results that one obtains for $d=3$ and $\gamma=\frac{1}{2}$ or $\gamma = \frac{3}{2}$. First of all, the analogue of relation (\ref{1stint}) in these cases yields,
\begin{eqnarray}
&& - \left[\beta \right]  = \frac{1}{16 r_{\rm s}} \left(\sqrt{2} r_{\rm s} \alpha - M \right)^2, \quad \gamma= \frac{1}{2}  , \label{betagamma12} \\
&& - \left[\beta \right] = \frac{4 M^2 r_{\rm s}}{\left( r_{\rm s} \pm \sqrt{2} M \alpha \right)^2}  , \quad \gamma = \frac{3}{2} . \label{betagamma32}
\end{eqnarray}
Using these relations, one can obtain the corresponding effective potentials. We find that for $\gamma= \frac{1}{2}$, oscillatory configurations exist in the $\alpha <0$ branch; while, for $\gamma= \frac{3}{2}$, they exist on both $\alpha > 0$ and $\alpha <0$ branches, depending on the choice of the sign in the denominator of (\ref{betagamma32}).

Let us now briefly comment on the holographic interpretation. Collapsing shells correspond to states in the dual field theory that thermalise. However, it is important to note that the very concept of thermalisation is often observable dependent, see \emph{e.g.}~the discussions in \cite{Balasubramanian:2010ce, Balasubramanian:2011ur}. In  systems that do thermalise, correlations over arbitrarily long distances eventually settle to the corresponding thermal values. Thus, the dynamics {\it terminates} at a particular {\it thermalisation time}, depending on the energy-scale at which one is probing.

The oscillatory configurations, in comparison, are quite unique.  Let us say that for the given set of parameters the dynamics of the shell is confined in the radial range $r=r^-_{\rm s}$ to $r=r^+_{\rm s}$. Then, local observables, such as the expectation value of energy-momentum tensor, do not exhibit any imprint of the oscillatory dynamics, and hence are indistinguishable from usual thermal states.  For any non-local boundary operator that probes bulk region $r > r^+_{\rm s}$, the system is always static and thermal. For any non-local boundary operator that probes beyond this bulk region, the system never thermalizes. We numerically study spacelike geodesics in the next subsection and demonstrate this explicitly. Thus, we have a dynamical state, for which the thermalization time is either $t_{\rm therm}=0$ or $t_{\rm therm}= \infty$. Presently, we do not have a good understanding of the nature of this state in the dual field theory.

\subsection{Geodesics in oscillating shells}
\label{geodesics}

To probe the oscillatory dynamics from the perspective of the boundary theory, we compute the equal time two-point function of an operator of large conformal dimension in the geodesic approximation. Such a  two-point  function  via a saddle point approximation~\cite{Banks:1998dd, Balasubramanian:1999zv} is: \be \langle \cO(\vec x) \cO(\vec x') \rangle \sim e^{- 2 \Delta \Sigma \left(\vec x, \vec x' \right) }.\ee Here $\cO$ and $\Delta$ are the operator and its conformal dimension, respectively. The length of the bulk spacelike geodesic  connecting the points $(t,\vec x)$ and $(t, \vec x')$ is denoted by $\Sigma \left(\vec x, \vec x' \right)$. Our goal here is to capture the imprint of the oscillatory dynamics on this correlation function.

We  study  geodesic lengths of spacelike geodesics anchored at a fixed value of boundary angular separation, $\Delta\varphi$. Since the shell expands and contracts periodically, the geodesics experience varying conditions near the shell.  This is expected to lead to an oscillatory evolution of the correlation function, which we verify by an explicit calculation.  To calculate the geodesics we follow~\cite{Taanila:2015sda}. A geodesic anchored at two points at the same time on the boundary must have a turning point in the bulk. The turning point is characterized by vanishing of the radial and temporal derivatives with respect to the proper length of the geodesic.

We impose these boundary conditions in the bulk at the turning point $(\bar{t},\bar{r})$.  Then we integrate the geodesic equations towards the AdS boundary. The data at the turning point map to the data at the boundary. The  data we need to extract from such geodesics include the angular separation, time at the boundary, and the geodesic length, denoted $\Delta \varphi_{\rm b}, t_{\rm b}, \Sigma$. By varying $(\bar{t},\bar{r})$ in the bulk and solving the geodesic equations, we generate a boundary dataset $\left(\Delta\varphi_{\rm b}, t_{\rm b}, \Sigma \right)$.

The affinely parameterised spacelike geodesic equations can be easily integrated both in the inside and outside regions to give the following first order equations: 
 \begin{align}
f_{\pm}t'~ =&~ E_{\pm}  \label{teqn},  \\
r^2\phi' ~=&~  L_{\pm}    \label{phieqn}	\ , \\
(r')^2 ~=& ~f_{\pm}\left( 1-\frac{L_{\pm}^2}{r^2} \right) + E^2_{\pm}.	\label{reqn}
 \end{align}
In these equations prime denotes derivatives with respect to the proper distance $\sigma$ along the spacelike geodesic.
Here $E_{\pm}$ and $L_{\pm}$ are the constants of motion. To ensure that the geodesic smoothly crosses the shell, we need to match the constants of motion appropriately on the two sides of the shell. To this end, we follow the treatment of~\cite{Keranen:2015fqa, Taanila:2015sda}.

\begin{figure}[ht!]
\begin{center}
{\includegraphics[width=0.45\textwidth]{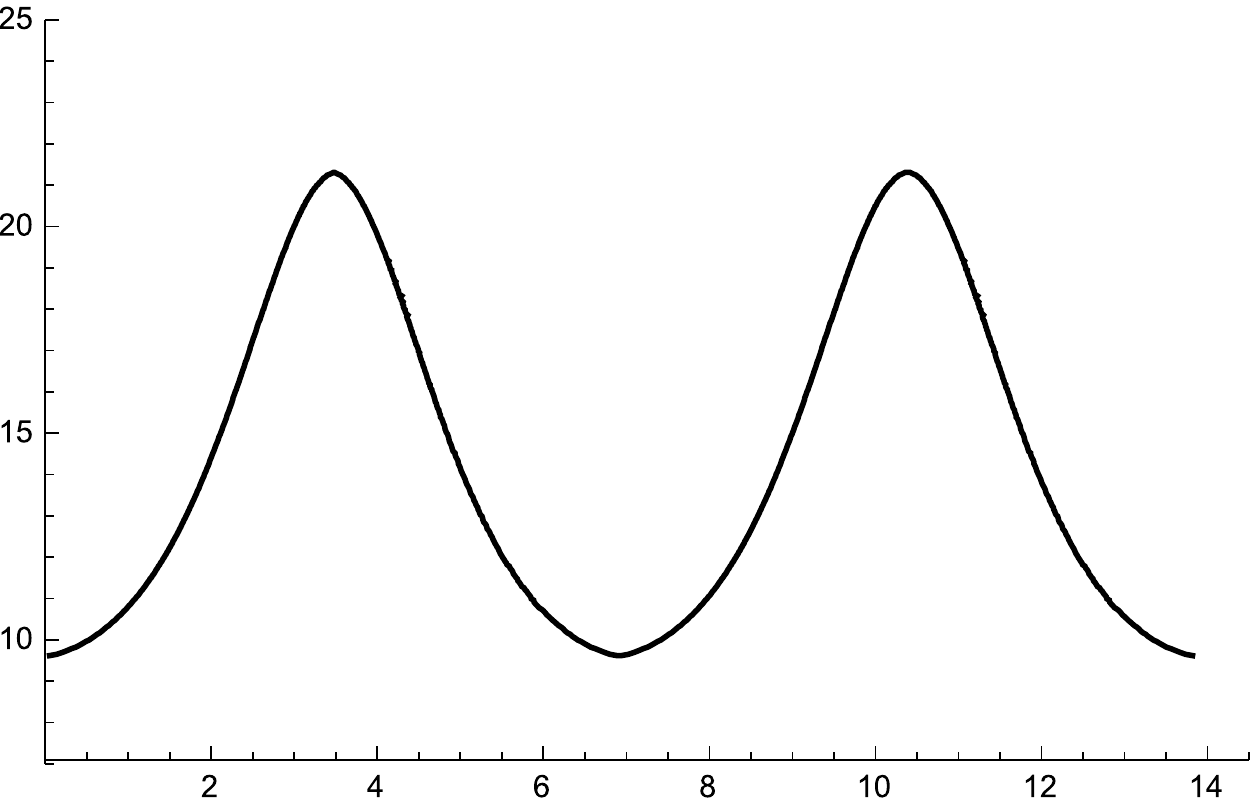}}
\caption{\sf A typical behaviour of renormalised geodesic length for fixed $\Delta \phi = 0.15$ as a function of time $t_+$ when the shell undergoes oscillatory motion. 
The coordinate $t_+$ is taken to be zero at the beginning of an oscillation cycle, when the shell is at its lower turning point.  Various parameters are: $\ell = 1, d = 4, \alpha = 0.992, m=24.45$, and the lower turning of the shell is taken to be at  $r^{-}_{\rm s} = 6.90$. The rest of the parameters are fixed by these values. In our conventions the $y$ axis is $10^{20}$ times $e^{-L}$ where $L$ is the proper length of the geodesics. } \label{oscgeod1}
\end{center}
\end{figure}

\begin{figure}[h!]
\begin{center}
{\includegraphics[width=0.45\textwidth]{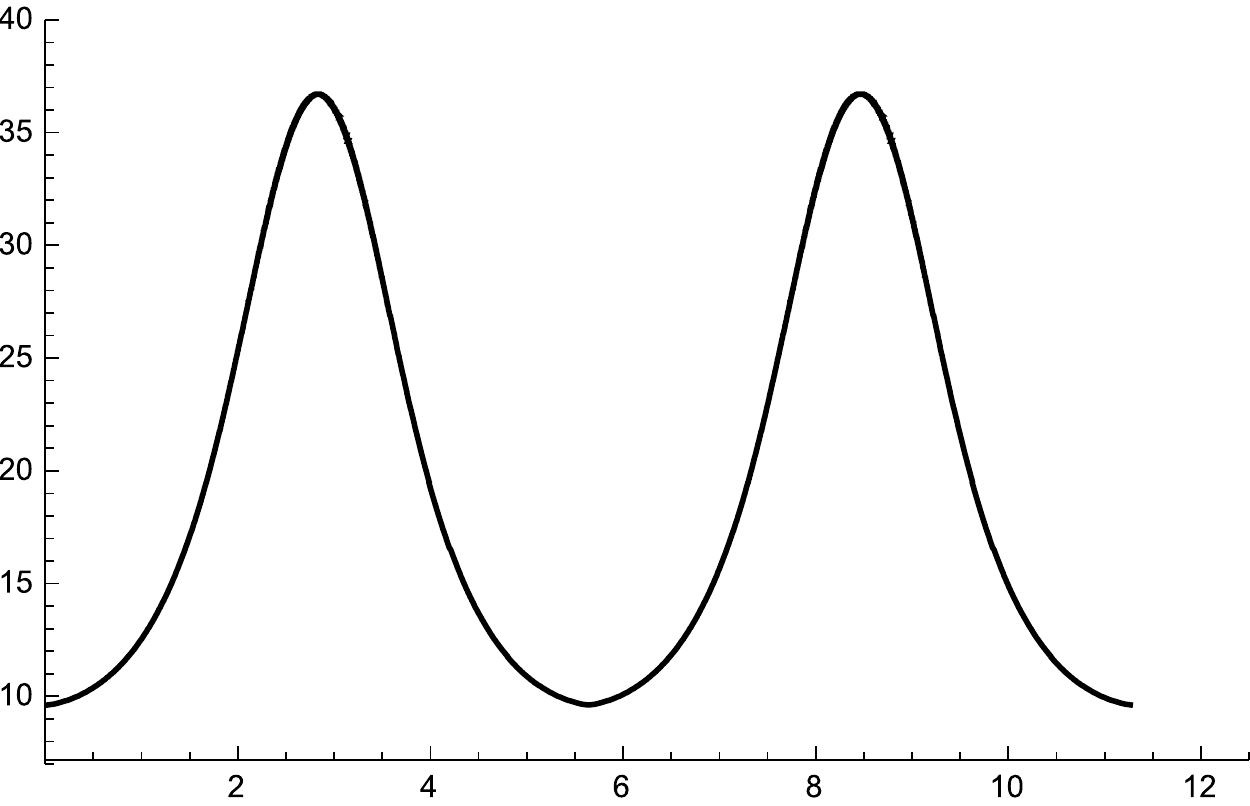}}
\caption{\sf Renormalised geodesic length for fixed $\Delta \phi = 0.15$ as a function time $t_+$ with different pressure on the shell. Parameter values are same as in Figure~\ref{oscgeod1} except $\alpha = 0.995$. For different values of $\alpha$, while keeping the other parameters same, the location of the upper turning point $r^+_{\rm s}$ of the shell  changes. As a result the oscillation period is different.  Geodesic lengths are also different.} \label{oscgeod2}
\end{center}
\end{figure}

The idea is to construct a coordinate system that is sufficiently smooth in a neighbourhood across the shell, and use it to transform quantities from the inside of the shell to the outside. The time coordinate for this coordinate system is chosen to be the proper time of the shell $\tau$. The spatial coordinate is naturally chosen to be the proper distance $\lambda$ away from the shell along spacelike geodesics normal to the shell. In terms of our inside and outside regions, the coordinate transformations,
 \begin{equation}
 (t_{\pm},r, \theta_i) \to (\tau,\lambda,\theta_i).		\label{coordtfmn}
 \end{equation}
 do the work. Using these coordinates one arrives at the  equations relating $t_-'$ to $t_+'$ and relating  $r'$ from the inside to the outside region~\cite{Keranen:2015fqa}\footnote{See section 2.2 and appendix B of~\cite{Keranen:2015fqa} for details.}
\begin{align}
\frac{dt_-}{d\sigma}\bigg\rvert_{r=r_{\rm s}}=&~~ \frac{dt_+}{d\sigma}\bigg\rvert_{r=r_{\rm s}}\frac{\beta_{{\rm s}-}\beta_{{\rm s}+}-\dot{r}_{\rm s}^2}{f_-} + \frac{dr_+}{d\sigma}\bigg\rvert_{r=r_{\rm s}} \ \frac{\dot{r}_{\rm s}}{f_-f_+}(\beta_{{\rm s}+}-\beta_{{\rm s}-}), 	\label{intext1} \\
\frac{dr_-}{d\sigma}\bigg\rvert_{r=r_{\rm s}} =&~~ \frac{dt_+}{d\sigma}\bigg\rvert_{r=r_{\rm s}}\dot{r}_{\rm s}(\beta_{{\rm s}+}-\beta_{{\rm s}-}) + \frac{dr_+}{d\sigma}\bigg\rvert_{r=r_{\rm s}}\frac{\beta_{{\rm s}-}\beta_{{\rm s}+}-\dot{r}_{\rm s}^2}{f_+} , \label{intext2}
\end{align}
where $\beta_{{\rm s}\pm}=\sqrt{f_{{\rm s}\pm}+\dot{r}_{\rm s}^2}.$ 
The inside and the outside derivatives of the continuous radial coordinate $r$ at the location of the shell are denoted as $\frac{dr_+}{d\sigma}\big\rvert_{r=r_{\rm s}}$ and $\frac{dr_-}{d\sigma}\big\rvert_{r=r_{\rm s}}$. Note that we only need to know the first derivatives of the $\pm$ coordinates with respect to the paramter $\sigma$ to match the geodesic across the shell. These conditions together with equations  (\ref{teqn})--(\ref{reqn})
allow us to relate $E_+$ to $E_-$.

We solve  equations  (\ref{teqn})--(\ref{reqn}) separately for the inside and the outside and match them across the shell according to (\ref{intext1}) and (\ref{intext2}). We begin by integrating a geodesic from its turning point $(\bar{t},\bar{r})$ in the inside region. At the turning point,  \bea t'&=&0, \\ r'&=&0,\eea which fixes the constants of motion to be,
\bea
E_-&=&0,\\
L_-&=&\bar{r}.
\eea
We integrate geodesic equations (\ref{teqn})--(\ref{reqn}) up to the location of the shell $r_{\rm s}$ with $f = f_-$. At this point, we switch to using the function $f_+$. We also need to use the constants of motion for the exterior. These are given by
\bea
E_+ &=& \sqrt{1-\frac{\bar{r}^2}{r_{\rm s}^2}}\frac{\dot{r}_{\rm s}}{\sqrt{f_-}}(\sqrt{f_++\dot{r}_{\rm s}^2}-\sqrt{f_-+\dot{r}_{\rm s}^2})  \label{eplus}
,  \\ 
L_+ &=& \bar{r}, \eea
where $E_+$ is deduced using (\ref{intext1}) or (\ref{intext2}) and the geodesic equations. Since the $\phi$ and the $r$ coordinates are continuous the conserved angular momentum does not change $L_+ = L_-$.

In figures~\ref{oscgeod1} and \ref{oscgeod2} we have plotted the geodesic lengths for a fixed value of boundary angular separation $\Delta\varphi$ as a function of the boundary time
$t_+$. The coordinate $t_+$ is taken to be zero at the beginning of an oscillation cycle where $r_{\rm s}=r^-_{\rm s}$. We can clearly see that the geodesic length oscillates with a fixed period. The period precisely corresponds to the period of the oscillation of the shell. Thus, the two-point function under study in the oscillating shell background  captures features of the oscillations.

We have chosen to plot geodesic lengths as a function of the time $t_+$. One can straightforwardly relate $t_+$ to the proper time of the shell $\tau$ or to $t_-$. We did not find any qualitative difference between the above graphs and the ones where the $x$-axis is taken to be proper time $\tau$ on the shell. We want to emphasise that our aim is to illustrate the qualitative behaviour of spacelike geodesics in oscillating shell backgrounds, as opposed to a detailed numerical analysis of these equations. At the  turning points $\dot r_{\rm s}$ vanishes, and naively there are 1/0 type expressions  encountered while doing numerical integrations. We  regulate such nuisances with a simple minded approach. For example, in the specific example of $1/\dot r_{\rm s}$, instead of taking the integration from $r=r^-_{\rm s}$ we take it from $r=r^-_{\rm s} + \epsilon$ with sufficiently small epsilon (and check that our results to do not depend on epsilon).

\section{The oscillating balls}

In this section we  consider the motion of a ball of matter of uniform density and pressure under its own gravity. The case of pressure-less dust was studied by Oppenheimer and Snyder~\cite{Oppenheimer:1939ue}.  In the context of the AdS/CFT correspondence, references \cite{Giddings:2001ii, Taanila:2015sda} studied similar dynamical situation in AdS background. We consider non-vanishing pressure. We are specifically  interested in exploring the possibility of oscillatory motion of the ball.

\label{osc_balls}

 \subsection{Oscillating FRW solutions}

The interior of a $d$-dimensional solid ball can be described by a Friedmann-Robertson-Walker (FRW) metric with $k = +1$, {\it i.e.}~positively curved $t = {\rm constant}$ slices,
\begin{equation}
ds_{-}^2= - dt^2 + R^2(t) \left( d\chi^2 + \sin^2\chi \ d\Omega_{d-1}^2 \right), \label{frwin}
\end{equation} 
sourced by perfect fluid stress-tensor
\begin{eqnarray}
T_{\mu \nu} = (\sigma + p)u_{\mu}u_{\nu} + p g_{\mu \nu}, \quad u^{\mu} = (1,0, \ldots ,0),
\end{eqnarray}
with an equation of state $p = w \sigma$. The radial and the time coordinates are denoted by $\chi$ and $t$, respectively. The function $R(t)$ is the scale factor. 

Einstein equations give the Friedmann equation for the scale factor
\begin{equation}
1 + R^2 + \dot{R}^2 = \frac{2\kappa \sigma}{d(d-1)} R^2, \label{Friedmann_eq_1}
\end{equation}
where we have used the value of the cosmological constant $\Lambda = - \frac{d (d-1)}{2\ell^2}$ and have set the AdS length $\ell$ to unity. The conservation of the energy-momentum tensor gives
\begin{equation}
\sigma R^{(w+1)d} = {\rm constant}. \label{conservation}
\end{equation}
Eliminating $\sigma$ from  the Friedmann equation \eqref{Friedmann_eq_1} using the  conservation equation \eqref{conservation} we get,
\begin{equation}
1 + R^2 + \dot{R}^2 = \frac{2\kappa \sigma_{0}R_0^{(w+1)d}}{d(d-1)R^{(w+1)d-2}},  \label{Friedmann_eq_2}
\end{equation}
where $\sigma_0$ is the initial density of the collapsing matter and $R_0$ is the initial scale factor.

We are interested in knowing if oscillatory solutions are possible to equation \eqref{Friedmann_eq_2}. In order to explore this, we rewrite that equation as
\be
\dot{R}^2 + V_{\rm eff}(R) = 0,
\ee 
with the effective potential
\be
V_{\rm eff}(R) = 1 + R^2 - \frac{c}{R^\beta},\label{ballpot}
\ee
where
\bea
c = \frac{2\kappa \sigma_{0}}{d(d-1)} R_0^{(w+1)d}, \qquad \qquad \beta =(w+1)d-2.
\eea

For oscillatory dynamics, the effective potential \eqref{ballpot} must develop a minimum in between two roots of equation $V_{\rm eff}(R) =0$. Let the roots be at $R=R_1$ and $R=R_2$ and the minimum be at $R= R_*$  with $R_1 < R_* < R_2$. Then, 
\be
V_{\rm eff}(R_*)<0, \qquad  V_{\rm eff}'(R_*)=0  \qquad \mbox{and} \qquad V_{\rm eff}''(R_*)>0.
\ee 
It is straightforward to see that
\begin{eqnarray}
V_{\rm eff}'(R_*) =0 \quad \implies \quad \beta = - \frac{2}{c} R_*^{2 + \beta}. \label{beta_eq}
\end{eqnarray}
For physically reasonable  initial parameters  $\sigma_0 >0$ and $R_0 >0$, thus the parameter $c$ is positive. Equation \eqref{beta_eq} then implies that $\beta <0$, \ie,
\be
w < - \left(\frac{d-2}{d}\right).
\ee

The second derivative of the potential \eqref{ballpot} at $R=R_*$ is
\begin{equation}
V_{\rm eff}''(R_*) = 2 \left(\beta + 2 \right).
\end{equation}
Requiring $V_{\rm eff}''(R_*) > 0$ gives $\beta > -2$ or equivalently $w > -1$. Thus, within the range
\be 
0 > \beta > -2, \qquad \qquad  -1 < w < - \left(\frac{d-2}{d}\right), \label{range_alpha}
\ee 
oscillatory ball dynamics is possible. Curiously the pressure $p = w \sigma$ must always be negative. We analyse the issue of energy conditions in section \ref{energy_conditions}. Next we comment on whether such an oscillatory FRW solution can be matched to an appropriate exterior solution.

 \subsection{Matching to an exterior star}
 
In the Oppenheimer-Snyder (OS) model the FRW metric that describes the interior of a collapsing star is matched to an empty Schwarzschild solution that describes the exterior of the collapsing star. The FRW metric is supported only by uniform pressure-less dust. The fact that such a smooth matching can be done is a remarkable fact about the OS model.
The pressure-less nature of the interior solution is an important ingredient.
 The OS model has been generalised to AdS space, see \emph{e.g.}~\cite{Giddings:2001ii}.

Here we are interested in  a generalisation of the OS model in AdS with non-zero pressure. In particular, we are interested in knowing if an oscillatory solution of the previous subsection can be taken to 
be the interior  of an  oscillating configuration in AdS. This turns out to be a difficult problem to analyse. 
In appendix \ref{OS_p_AdS} we report some progress on this problem. We construct a matched metric when the equation of state $p =p(\sigma)$ is arbitrary, and can be chosen  independently for the interior and the exterior of the model. The distinction between the interior and the exterior is as the two sides of a ``shock wave'' across which the metric is continuous.  We find that the pressure and energy density suffer a discontinuity across the shock surface. Such shock waves are the counterparts  of fluid dynamical shock waves on curved backgrounds. A detailed study of such systems was done by Smoller and Temple~\cite{SmollerTemple}, who also constructed a flat space generalisation of the OS model with non-zero pressure. Our analysis in the appendix closely follows their construction.  

When we demand that the extrinsic curvature also remains continuous (as in the OS model, and in contrast to the thin-shell model), the set-up becomes over-constrained. One way to achieve extrinsic curvature continuity is by {\emph{not}} demanding an equation of state for the interior or for the exterior  solution. We can treat pressure and density as independent dynamical variables, say for the interior solution. By doing so, one can fix the pressure and density for the interior solution from the exterior solution.   This strategy has its shortcomings, but this is one way in which interior and exterior solutions can be matched~\cite{SmollerTemple}. We illustrate how such a matching is to be done, from a given outside solution  to an appropriate inside solution.
For our problem, however,  the matching needs to done the other way, {\it i.e.}, given an FRW solution of the previous subsection, can we find an appropriate exterior solution? Unfortunately, we do not know a full answer to this question. Given the analysis of appendix \ref{OS_p_AdS}, it seems feasible that some exterior star solution can be matched to a given interior solution, however, the precise details of such an analysis are likely to be complicated and are left for future investigations.

\section{Energy conditions}
\label{energy_conditions}

In this section we analyse various energy conditions for the above discussed oscillating solutions. 

\subsection{Oscillating shells}

In the case of oscillating shells one can consider two independent notions of energy conditions. One is associated with the shell stress-energy tensor (\ref{shellmatter}) and the other is associated with the Einstein tensor constructed from the induced metric (\ref{shellinduced}). Interestingly, these two turn out to have independent characters, as we discuss below.

\subsubsection*{Energy conditions with $S_{ab}$}

Since the surface stress tensor (\ref{shellmatter}) is of the perfect fluid form with $\sigma$ and $p$ given by (\ref{israelexplicit}), null energy condition is equivalent to the statement that  $\sigma + p \geq 0 $ and the weak energy condition is equivalent to the statement that $\sigma \ge 0, \, \sigma + p \geq 0$. 

 At the turning points of the oscillating shell where $\dot{r}_{\rm s} = 0$, we have from equations (\ref{israelexplicit}),
\begin{eqnarray}
\sigma & =& \frac{d-1}{ \kappa r_{\rm s}} \bigg(\sqrt{1+ r_{\rm s}^2} - \sqrt{1+r_{\rm s}^2 - \frac{m}{r_{\rm s}^{d-2}}}\bigg), \label{sigma} \\
\sigma + p &=& \frac{ r_{\rm s} + \frac{(d-2)}{2}\frac{m}{r_{\rm s}^{d-1}} - \frac{V_{\rm eff}'(r_{\rm s})}{2}}{\kappa \sqrt{1+ r_{\rm s}^2-\frac{m}{r_{\rm s}^{d-2}} }}- \frac{r_{\rm s} - \frac{V_{\rm eff}'(r_{\rm s})}{2}}{\kappa \sqrt{1+ r_{\rm s}^2}}+ \frac{1}{ \kappa r_{\rm s}}\bigg(\sqrt{1+ r_{\rm s}^2} - \sqrt{1+ r_{\rm s}^2 - \frac{m}{r_{\rm s}^{d-2}}}\bigg). \label{sigmap}
\end{eqnarray}
From these expressions it is clear that $\sigma$ is positive definite, provided $m > 0$. The right hand side of expression (\ref{sigmap}) is positive definite provided $2 r_{\rm s} > V_{\rm eff}' (r_{\rm s})$. For a given configuration ({\it i.e.}, a given set of parameters), a straightforward numerical check can confirm if this is indeed the case or not. For the cases we have checked, we found that both weak and null energy conditions are satisfied at the turning points. 
We also found that for all the cases that we have checked, $\sigma +p $ is positive for the entire motion of the oscillatory shells. Thus, the null and weak energy conditions seem to be satisfied for the surface stress tensor all along the oscillation of the shell.

\subsubsection*{Energy conditions with $G_{ab}$}
From the induced metric 
 \eqref{shellinduced}
we can define an effective stress tensor $\kappa T_{ab} := G_{ab} = R_{ab} - \frac{1}{2} g_{ab} R$. This stress tensor turns out to be of the perfect fluid form, which allows us to define an effective energy density and pressure. We find
\bea
\sigma_{\rm eff} =  \frac{(d-1)(d-2)}{2 \kappa  r_{\rm s}^2}(1+ \dot{r}_{\rm s}^2 ),   \qquad \qquad p_{\rm eff} = - \frac{(d-2)}{\kappa} \left( \frac{\ddot{r}_{\rm s}}{r_{\rm s}} 
+ \frac{(d-3)(1+ \dot{r}_{\rm s}^2)}{2r_{\rm s}^2}\right),
\eea
as a result
\be
\sigma_{\rm eff} + p_{\rm eff} = \frac{(d-2)}{\kappa r_{\rm s}^2}(1+ \dot{r}_{\rm s}^2  - r_{\rm s}\ddot{r}_{\rm s}).
\ee

At the bounce $\dot r_{\rm s}^2 =0$, therefore
 \be
\sigma_{\rm eff} + p_{\rm eff} =  \frac{(d-2)}{\kappa r_{\rm s}^2}(1- r_{\rm s}\ddot{r}_{\rm s}).
\ee
We note that $\sigma_{\rm eff} + p_{\rm eff} > 0$ at the bounce provided $\ddot{r}_{\rm s} < r_{\rm s}^{-1}$, {\it i.e.}, if the bounce is sufficiently `gentle'. A very similar  set of conditions were discussed in \cite{MolinaParis:1998tx} in a different context. From the definition of effective potential \eqref{shelldyn1}, we have  $\ddot{r}_{\rm s} = - \frac{1}{2} V_{\rm eff}'(r_{\rm s})$. Therefore,  
\be
\sigma_{\rm eff} =  \frac{(d-1)(d-2)}{2 \kappa  r_{\rm s}^2}(1 - V_{\rm eff}(r_{\rm s})), \qquad \qquad  \sigma_{\rm eff} + p_{\rm eff} = \frac{(d-2)}{\kappa r_{\rm s}^2}\left(1 - V_{\rm eff}(r_{\rm s})+  \frac{1}{2}r_{\rm s}V_{\rm eff}'(r_{\rm s}) \right).
\label{riccishell}
\ee%
The energy density $\sigma_{\rm eff}$  so defined is always positive. For a given set of parameters,  one can easily check  numerically whether $\sigma_{\rm eff}+p_{\rm eff}$ is positive or not. We find that for all the cases that we have checked $\sigma_{\rm eff}+p_{\rm eff} > 0$ for oscillatory shells. Therefore, null and weak energy conditions so defined also seem to be satisfied all along the motion of the shell.

\subsection{Oscillating balls}
Now we discuss the energy conditions for the oscillating FRW metrics of section \ref{osc_balls}. The energy density and pressure can be read off from the Einstein's equations. We get
\be
\sigma = \frac{d(d-1)}{2\kappa R^2} \left(1 + R^2 + \dot R^2\right), \qquad  \qquad \sigma + p = \frac{(d-1)}{\kappa R^2} (1+ \dot{R}^2  - R\ddot{R}).
\ee
Using the effective potential we once again get expressions very similar to \eqref{riccishell},
\be
\sigma= \frac{d(d-1)}{2\kappa R^2} (1 + R^2 - V_{\rm eff}(R)), \qquad \qquad  \sigma+p = \frac{(d-1)}{\kappa R^2} \left(1 - V_{\rm eff}(R)+  \frac{1}{2}RV_{\rm eff}'(R) \right).
\label{ball_energy_pressure}
\ee%
These expressions for the potential \eqref{ballpot} become, 
\be
\sigma= \frac{d(d-1)c}{2\kappa R^{\beta +2}}, \qquad \qquad  \sigma+p =\frac{(d-1)(\beta + 2)c}{\kappa R^{\beta + 2}} .
\ee
A requirement for oscillations is precisely $\beta + 2> 0$, cf.~\eqref{range_alpha}. Therefore, we note that both null and weak energy conditions are satisfied all along the motion of the ball. It is intriguing (and perhaps counterintuitive) that negative pressure is needed to sustain these oscillations. We note that with zero cosmological constant,  accelerated expansion also requires negative pressure precisely in the range \eqref{range_alpha}.\footnote{We thank Jorge Rocha and Vitor Cardoso for this observation.}

\subsection*{Acknowledgements}
It is a pleasure to thank Sudipta Mukherji for discussions and for suggesting us to explore energy conditions. The work of AV is supported in part by the DST-Max Planck Partner Group ``Quantum Black Holes'' between IOP, Bhubaneshwar and AEI, Golm. AV and PR thank AEI Golm for warm hospitality where part of this work was done. AB thanks IOP Bhubaneshwar for warm hospitality where part of this work was done.


\appendix 


\section{Oppenheimer-Snyder model with non-zero pressure in AdS}

\label{OS_p_AdS}
The OS model requires that the pressure of the collapsing star be identically zero. 
In this appendix we construct a generalisation of the Oppenheimer-Snyder (OS) model in AdS with non-zero pressure. We first construct a matched metric when the equation of state $p =p(\sigma)$ is arbitrary, and can be chosen  independently for the interior and the exterior of the model. The distinction between the interior and the exterior is as the two sides of a ``shock wave'' across which the metric is continuous. 
When we demand that the extrinsic curvature also remains continuous, the set-up becomes over-constrained. One way to achieve extrinsic curvature continuity is by {\emph{not}} demanding an equation of state for the interior Friedmann-Robertson-Walker (FRW) solution. That is, to treat pressure and density as independent dynamical variables. Doing this has its shortcoming, but this is one way in which exterior solution can be matched to an interior solution with pressure.\footnote{Another approach could be to introduce a surface stress-tensor at the interface, as in the thin-shell model. We do not pursue this idea here.}

We find that the pressure and energy density suffer a discontinuity across the shock surface. Such ``shock waves'' are the counterparts  of fluid dynamical shock waves on curved backgrounds. A detailed study of such systems was done by Smoller and Temple \cite{SmollerTemple}, who also constructed a flat space generalisation ({\it i.e.},~with $\Lambda = 0$) of the OS model with non-zero pressure. Our analysis below closely follows their construction. We work in four spacetime dimensions. This appendix is a preliminary study, it serves to illustrate how such a matching is to be done, from a given outside solution  to an appropriate inside solution. We do not address several physics issues {\it e.g.} energy conditions for the inside solution,  or if the matching can be done the other way -- given an inside solution can it be matched to an appropriate outside solution?

\subsection*{Interior solution: FRW in AdS}
As in the OS model, the inside metric is taken to be the Friedmann-Robertson-Walker (FRW) solution
\be
ds^2 = - dt^2 + R^2(t) \left(\frac{1}{1-kr^2}dr^2 + r^2 d \Omega^2\right), \label{FRWmetric}
\ee
with the perfect fluid stress tensor source where $p$ and $\sigma$ only depend on time $t$. Einstein's equations give
\bea
\dot \sigma &=& - 3\frac{\dot R}{R} (p + \sigma),  \label{FRWe1}\\
\dot R^2 + k &=& \frac{8\pi}{3} \sigma R^2 - \frac{R^2}{\ell^2} \label{FRWe2},\\
\frac{\ddot R}{R} &=& - \frac{4\pi}{3} (\sigma + 3 p) - \frac{1}{\ell^2}.
\eea
Equation  \eqref{FRWe1} is equivalent to  
\be
\frac{d}{dR} (\sigma R^3) = - 3 p R^2.
\ee
When pressure is zero this equation tells that the ``mass'' $M = \frac{4\pi}{3}\sigma R^3$ contained inside the star remains constant as the star evolves in time. With non-zero pressure we see that this is not the case. There is exchange of matter between the interior and the exterior, which needs to be carefully taken into account while matching the two solutions.

\subsection*{Exterior solution: TOV equations in AdS}

The non-zero pressure for the interior solution also requires non-zero pressure for the exterior solution. This is so, because in the presence of pressure, matter can flow across the shock surface. Thus the exterior geometry is a spherically symmetric ``star'' with non-zero pressure and energy density, as opposed to the vacuum Schwarzschild solution in the OS model. Such configurations are described by the Tolman-Oppenheimer-Volkoff (TOV) equations.  Therefore, our next  aim is to get the TOV equations in AdS space. Let us start with the metric (see also \cite{deBoer:2009wk})
\be
d\bar s^2 = - B(\bar r) d\bar t^2 + A(\bar r)^{-1} d \bar r^2 + \bar r ^2 d \Omega^2.
\label{TOVmetric}
\ee
Typically this set-up is used for describing the ``interior of a star'' but in our case it describes the ``exterior of the ball''. It is supported by the perfect fluid stress tensor
\be
\bar T_{\mu \nu} = \bar p g_{\mu \nu} + (\bar p + \bar \sigma) \bar u_{\mu} \bar u_{\nu},
\ee
with some equation of state 
\be
\bar p = \bar p (\bar \sigma),
\ee
and where fluid is taken to be not moving, {\it i.e.}, 
\be
\bar u_{\mu} = (\sqrt{B(\bar r)},0,0,0).
\ee

The above equations are all written in barred notation so that they can be distinguished from the interior unbarred notation when we do the matching. 
We take the function $A(\bar r)$ to be of the form
\be
A(\bar r) = \left( 1- \frac{2M(\bar r)}{\bar r } + \frac{\bar r^2}{\ell^2}\right),
\ee
where we have set Newton's constant and the speed of light to unity, but the AdS length is kept explicitly for clarity. The function $M(\bar r)$ is so far undetermined. It is akin to the ADM mass. Einstein's equations in four-dimensions,
\be
R_{\mu \nu} - \frac{1}{2} R g_{\mu \nu} - \frac{3}{\ell^2} g_{\mu \nu} = 8 \pi \bar T_{\mu \nu},
\ee
give the following ordinary differential equations,
\begin{align}
\frac{dM(\bar r) }{d \bar r}&= 4 \pi \bar r^2 \bar \sigma, \label{TOV1} \\
\frac{B'(\bar r)}{B(\bar r)}&= -  \frac{2\bar p'(\bar r)}{ \bar p + \bar \sigma},  \label{TOV2} \\
- \bar r^2 \frac{d}{d \bar r} \bar p &= M(\bar r) \bar \sigma \left(1 + \frac{\bar p}{\bar \sigma}\right) \left(1 + \frac{\bar r^3}{M(\bar r)}\left( 4 \pi \bar p + \frac{1}{\ell^2} \right) \right) \left( 1- \frac{2M}{r} + \frac{r^2}{\ell^2}\right)^{-1}. \label{TOV3}
\end{align}
These equations are the generalisation of text book TOV equations with non-zero cosmological constant. In the limit $\ell \to \infty$ they reduce to the standard TOV  equations, see {\it e.g.}~\cite{Weinberg}.

A slightly better presentation is possible if we work with the following variables  \cite{deBoer:2009wk}
\be
B(\bar r) = A(\bar r) e^{2 \chi(\bar r)}.
\ee
Then the above equations simplify to 
\bea
\frac{dM(\bar r) }{d \bar r}&=& 4 \pi \bar r^2 \bar \sigma, \\
p'(\bar r) &=& -  \frac{1}{2}\frac{B'(\bar r)}{B(\bar r)}( \bar p + \bar \sigma),\\
\chi'(\bar r) &=& 4 \pi \bar r (\bar p + \bar \sigma) A(\bar r)^{-1}.
\eea

\subsection*{Matching}
We now do the matching and also find the matching surface. In order to do so, we construct $(\bar t, \bar r)$ coordinate system for the FRW metric. To make sure that the areas of 2-spheres agree in the two coordinate systems at the matching surface, we must demand
\be
\bar r = R(t) r.
\label{match1}
\ee
We first write FRW metric in $(t, \bar r )$ coordinates.
From \eqref{match1} we have
\be
d\bar r = R dr + \dot R r dt.
\ee
Using this, the FRW metric \eqref{FRWmetric} can be written in the $(t, \bar r )$ coordinates as
\be
ds^2 = - \left\{1 - \frac{\dot R^2 \bar r^2}{R^2 - k \bar r^2} \right\} dt^2 + \frac{R^2}{R^2 - k \bar r^2} d \bar r^2 - \frac{2 R \dot R \bar r}{R^2 - k \bar r^2} dt d \bar r + \bar r^2 d\Omega^2,
\ee
which upon inserting \eqref{FRWe2} becomes
\be
ds^2 =\frac{1}{R^2  - k \bar r^2} \left\{ - R^2 \left( 1 - \frac{8\pi}{3} \sigma R^2  r^2 + \frac{ r^2 R^2}{\ell^2}\right) dt^2 + R^2 d \bar r ^2 - 2 R \dot R  \bar r d\bar r dt \right\} + \bar r ^2 d\Omega^2.
\label{metrictbarr}
\ee

Our aim is to match the interior metric in $(\bar t, \bar r)$ coordinates to the TOV metric \eqref{TOVmetric}. TOV metric does not have any cross-term, therefore, we next define a mapping $t = t(\bar t, \bar r)$ to eliminate the cross term $d\bar r dt$ in  metric \eqref{metrictbarr}. 

It is notationally more convenient to  consider the general metric of the form
\be
d\tilde s^2 = - C(t,\bar r) dt^2 + D(t, \bar r) d\bar r^2 + 2 E(t,\bar r) dt d \bar r.
\label{ds2tilde}
\ee
Consider a function $\psi(t, \bar r)$ that satisfies
\be
\partial_{\bar r} (\psi C) + \partial_t (\psi E) = 0. \label{PDE}
\ee
The coordinate $\bar t$ defined via
\be
d\bar t = \psi (C dt - E d \bar r) \label{tbart}
\ee
is an exact differential and also  eliminates the cross term in \eqref{ds2tilde} to give
\be
d\tilde s^2 = - (\psi^{-2} C^{-1}) d\bar t^2 + \left( D + \frac{E^2}{C} \right) d \bar r ^2.
\ee

Applying this recipe to metric \eqref{metrictbarr} and comparing $d \bar r ^2$ term with TOV metric  \eqref{TOVmetric}, we obtain the equation of the \emph{shock surface}
\be
M(\bar r) = \frac{4 \pi}{3} \sigma(t) \bar r^3. \label{surface}
\ee
This is an equation in the $(t, r)$ coordinates, since $\bar r = R(t) r$.  

The function $\psi$ needs to be determined such that $d\bar t^2$ terms from the two sides also match on the shock surface \eqref{surface}. This leads to the requirement
\be
\frac{1}{\psi^{2} R^{2}} \frac{1}{(R^2  - k \bar r^2)} \left( 1 - \frac{8\pi}{3} \sigma \bar r^2 + \frac{ \bar r^2}{\ell^2}\right)^{-1} = B(\bar r) \label{data}
\ee
on the shock surface.

The picture is as follows:  the function $\psi(t, \bar r)$ is determined by the solution of the first order linear partial differential equation \eqref{PDE} where
\bea
C&=& \left( 1 - \frac{8\pi}{3} \sigma \bar r^2 + \frac{ \bar r^2}{\ell^2}\right)R^2,\\
E&=&- R \dot R \bar r ,
\eea
subject to the initial data \eqref{data} on the surface \eqref{surface}. If this problem can be solved, the two metrics can be matched continuously.

\subsection*{Jump in density}
From equation \eqref{TOV1}, we have that the mass function $M(\bar r)$ for the TOV metric is given by
\be
M(\bar r_0) = \int_0^{\bar r_0} 4 \pi \bar \sigma(\bar r) \bar r^2 d \bar r.
\ee
In writing this equation we are imagining that the TOV metric is continued to $\bar r$  values less than that of the shock  surface. The quantity $M(\bar r)$ represents the total mass that is generating the TOV solution outside the shock wave.
For a physically reasonable model of a star
$
\frac{d\sigma}{d\bar r} <0,
$ 
therefore, 
\be
M(\bar r_0) > \frac{4 \pi}{3} \bar \sigma (\bar r_0) \bar r_0^3.
\ee
Compare this equation with \eqref{surface}. This allows us to conclude that at the shock surface
\be
\sigma > \bar \sigma, \qquad \qquad [\sigma] \equiv \bar \sigma - \sigma < 0,
\ee
{\it i.e.}, density inside is greater than the density outside.

\subsection*{Shock speed}
Differentiating \eqref{surface} with respect to $t$, we find the shock speed
\be
\dot{\bar{r}} = \frac{\dot \sigma \bar r}{3 [\sigma]}. \label{speed}
\ee
Since $[\sigma] <0$, the shock speed is negative if $\dot \sigma > 0$. We also note that $\dot \sigma $ is indeed positive for a collapsing situation as $\dot R < 0$, cf.~\eqref{FRWe1}.

\subsection*{Continuity of extrinsic curvature}
Smoller and Temple \cite{SmollerTemple} also show that in the present set-up, the continuity of the extrinsic curvature is equivalent to the statement that 
the normal-normal component of the external stress-tensor has no jump, 
\be
[T]^{\mu \nu} n_\mu n_\nu = 0. \label{jump}
\ee

Explicitly, we have for the inside
\bea
T^{\mu \nu} n_\mu n_\nu &=& p (n \cdot n) + (p + \sigma) (u \cdot n)^2, \\
&=& p (n \cdot n) + (p + \sigma) n_0^2,
\eea
where we have used the fact that $u^\mu = (1,0,0,0)$ for the FRW set-up. Similarly, we have  for the outside
\bea
\bar T^{\mu \nu} \bar n_\mu \bar n_\nu &=& \bar p (\bar n \cdot \bar n) + (\bar p + \bar \sigma) (\bar u \cdot \bar n)^2, \\
&=& \bar p ( \bar n \cdot \bar n) + \frac{1}{B(\bar r)}(\bar p + \bar \sigma) \bar n_0^2.
\eea
Therefore the jump condition \eqref{jump} becomes
\be
\bar p  (\bar n \cdot \bar n) - p (n \cdot n)  + \frac{1}{B(\bar r)} (\bar \sigma + \bar p) \bar n_0^2 -  (\sigma + p)n_0^2 =0. \label{jump2}
\ee
We note that $n_\mu$ and $\bar n_\mu$ are components of the same vector $n_\mu$. More explicitly, we write the shock surface as
\be
\varphi(t,r) = r - r (t)  =0.
\ee
with the normal $d\varphi = n_\mu d x^\mu$. This gives $n_0 = - \dot r$. To obtain components in the barred coordinates, we rewrite the shock surface as
\be
\varphi(\bar t,\bar r) = \frac{\bar r}{R(t(\bar t, \bar r))} - r (t(\bar t, \bar r))  =0,
\ee
which gives $\bar n_0 = - \frac{\dot{\bar r} }{R} \frac{\partial t}{\partial \bar t}$, where we have used the fact that $\bar r = r R(t)$. Equation \eqref{tbart} then yields,
\be
\bar n_0 = - \frac{\dot{\bar r} }{\psi C R}. \label{n0bar}
\ee
Inserting equation \eqref{data} into  expression \eqref{n0bar} gives 
\be
\bar n_0^2 = \frac{B}{AR^2}(1-kr^2)  \dot{\bar r}^2.
\ee
Using these various elements, the jump condition  \eqref{jump2} becomes,
\be
(\sigma + \bar p) \dot r^2 - (\bar \sigma + \bar p) \frac{1-kr^2}{AR^2}  \dot{\bar{r}}^2 + (p - \bar p) \frac{1-k r^2}{R^2} = 0. \label{jump3}
\ee

This equation is an additional constraint that must be satisfied on the shock surface. It is a complicated relation between $p, \bar p, \sigma, \bar \sigma, R$ on the shock surface $r = r(t)$. In the OS limit, where $\bar \sigma = \bar p = 0$ it reduces to 
\be
 \sigma \dot r^2  + p \frac{1-k r^2}{R^2} = 0.
 \ee
Under the assumption that $g_{rr}$ of the FRW metric is positive, {\it i.e.}, $\frac{1-k r^2}{R^2} > 0$ and $\sigma > 0$, we conclude that the only way this constraint can be satisfied is when
\bea
p &=& 0, \\
\dot r &=& 0.
\eea
This means that the FRW interior must be pressure free and the shock surface is $r= \verb+constant+$, which are both features of the OS model.

\subsection*{Inside solution, given the outside}
Now it seems that we have an over-constrained situation. Given  an equation of state $\bar p = \bar p(\bar \sigma)$, we can in principle integrate TOV equations to find the exterior solution on and outside the shock surface. For this solution to be matched to an interior FRW solution, we need to know $R(t), \sigma(t)$, and $p(t)$. Given an equation of state for the interior solution, we need to know only two functions, say, $R(t), \sigma(t)$. These two functions can be determined by the two Friedmann equations \eqref{FRWe1}--\eqref{FRWe2}. Then, how to ensure that the constraint \eqref{jump3} is satisfied? It seems that we have three equations for two variables.

The picture that Smoller and Temple proposed for this problem is to view pressure in the FRW metric as an independent dynamical variable, rather than fixed by an equation of state. The idea then is to determine $p$ from equation \eqref{FRWe1}. Substituting this $p$ in \eqref{jump3} to get an equation only involving $\sigma(t)$ and $R(t)$. Solution of that equation together with \eqref{FRWe2} completely specifies the FRW metric inside. 

More explicitly, it proceeds as follows. Rewriting \eqref{FRWe1}, we have
\be
p = -\sigma - \frac{\dot \sigma R}{3 \dot R}.
\ee
Using \eqref{speed} into this equation we get
\be
p = - \bar \sigma - [\sigma]\frac{R \dot r}{r \dot R}, \label{pequation}
\ee
which gives the variable $p(t)$ in terms of the unknowns $R(t)$ and $\sigma(t)$ on the shock surface $r(t)$. Substituting \eqref{pequation} in \eqref{jump3} gives the constraint equation \eqref{jump} in its most useful form,
\be
\alpha \dot r^2 + \beta \dot r + \gamma = 0,  \label{jumpmain}
\ee
with
\bea
\alpha &=& \frac{\sigma + \bar p}{1-kr^2} - \frac{\bar \sigma + \bar p}{A},\\
\beta &=&- \frac{2\dot R r}{A R} (\bar \sigma + \bar p) + \frac{1}{\bar r \dot R} (\sigma - \bar \sigma),\\
\gamma &=& - \left(1 + \frac{\dot R^2 r^2}{A} \right)\left(\frac{\bar \sigma + \bar p}{R^2}\right).
\eea

All functions appearing in \eqref{jumpmain} and \eqref{FRWe2} are expressed in terms of unknowns $r(t)$\footnote{Equivalently, $\sigma(t)$, cf.~\eqref{surface}.} and $R(t)$. The solution to these equations determines the shock surface and FRW scale factor, and from these two quantities we know $\sigma(t)$ via \eqref{surface} and  $p(t)$ via \eqref{pequation}. Hence the full interior FRW metric is determined. The matched FRW solution is such that the metric and the extrinsic curvature are continuous across the shock. 

\subsection*{Shortcomings}
As mentioned in the beginning of this discussion, {\it a priori} the above analysis does not ensure any physical condition for the interior solution. Since $p(t)$ and $\sigma(t)$ are explicitly known at the end of the procedure, one can always check if it is physically reasonable or not, {\it i.e.}, whether some energy condition is satisfied or not. Moreover, the way this construction is set-up, it allows us to match a given exterior TOV solution to an appropriate interior FRW solution. It is not at all obvious if the logic can be implemented  the other way round, namely, given an FRW solution (possible oscillating), can one find an appropriate TOV solution where the metric and  extrinsic curvature are  matched continuously? We leave this investigation for future studies.

\section{The effective potential}

In this section we explicitly write down the effective potential that we have studied throughout the paper, for the purpose of reproducibility for the interested readers. The junction conditions in (\ref{extrinsic}), (\ref{shellmatter}), along with the definition of $\beta_{\pm}$ in (\ref{tdotshell}) can be rewritten in the following form:
\begin{eqnarray}
&& \dot{r}_{\rm s}^2 + V_{\rm eff} \left(r_{\rm s} \right) = 0 \ , \\
&& V_{\rm eff} \left(r_{\rm s} \right) = f_-\left( r_{\rm s} \right) - \frac{\kappa^2}{4 \left[\beta \right]^2 } \left(f_-\left( r_{\rm s} \right) - f_+\left( r_{\rm s} \right) + \frac{\left[ \beta\right]^2}{\kappa^2} \right)^2 \ .
\end{eqnarray}
We consider a general polytropic equation of state of the form:
\begin{eqnarray}
p = \frac{\alpha}{d-1} \sigma^\gamma \ ,
\end{eqnarray}
where $p$, $\sigma$ are the pressure and energy density, respectively. The explicit expression for $\left[ \beta \right]$, for integer values of $\gamma \not = 1$, is given by
\begin{eqnarray}
- \left[ \beta \right]  = \left[ - \frac{ \left( d-1\right)^{\gamma -2}}{\kappa^{\gamma -1 }}\ \alpha \ r_{\rm s}^{1 - \gamma  } + M r_{\rm s}^{\left( d-2 \right) \left( \gamma -1 \right) } \right]^{\frac{1}{1-\gamma}} \ ,
\end{eqnarray}
in $(d+1)$-bulk dimensions, with an integration constant $M$. As we have also remarked in the main text, it is possible to obtain the expression for $\left[ \beta \right]$ for non-integer values of $\gamma$ as well, such as the one written in (\ref{betagamma12}), (\ref{betagamma32}). Finally, the two functions $f_{\pm}$ are given by
\begin{eqnarray}
f_- \left( r \right) =  1 + r^2 \ , \quad f_+ \left( r \right) =  1 + r^2 - \frac{m}{r^{d-2}} \ .
\end{eqnarray}
Now, with various values of the parameters in the problem, one can proceed to obtain and analyze the various properties of the corresponding effective potential.

\end{document}